\documentclass[preprint]{aastex631}

\newcommand{\dioxi}{CO\textsubscript{2} }
\newcommand{\water}{H\textsubscript{2}O }
\newcommand{\kb}{\texttt{KB/SD} }
\newcommand{\ooc}{\texttt{OC} }

\graphicspath{{./}{figures/}}

\begin{document}

\title{On the early thermal processing of planetesimals during and after the giant planet instability}



\author[0000-0002-1611-0381]{Anastasios Gkotsinas}
\affiliation{Laboratoire de G\'{e}ologie de Lyon: Terre, Plan\`{etes}, Environnnement, CNRS, UCBL, ENSL, F-69622, Villeurbanne, France}

\author[0000-0002-4547-4301]{David Nesvorn\'{y}}
\affiliation{Department of Space Studies, Southwest Research Institute, 1050 Walnut Street, Suite 300, Boulder, CO 80302, USA}

\author[0000-0003-2354-0766]{Aur\'{e}lie Guilbert-Lepoutre}
\affiliation{Laboratoire de G\'{e}ologie de Lyon: Terre, Plan\`{etes}, Environnnement, CNRS, UCBL, ENSL, F-69622, Villeurbanne, France}

\author[0000-0001-8974-0758]{Sean N. Raymond}
\affiliation{Laboratoire d'Astrophysique de Bordeaux, Univ. Bordeaux, CNRS, F-33615 Pessac, France}

\author[0000-0001-5272-5888]{Nathan Kaib}
\affiliation{Department of Physics and Astronomy, University of Oklahoma, 440 W. Brooks St., Norman, OK 73019, USA}

\begin{abstract}

Born as ice-rich planetesimals, cometary nuclei were gravitationally scattered onto their current orbits in the Kuiper Belt and the Oort Cloud during the giant planets' dynamical instability in the early stages of our Solar System's history. Here, we model the thermal evolution of planetesimals during and after the giant planet instability. We couple an adapted thermal evolution model to orbital trajectories provided by \textit{N}-body simulations to account for the planetesimals' orbital evolution, a parameter so far neglected by previous thermal evolution studies. Our simulations demonstrate intense thermal processing in all planetesimal populations, concerning mainly the hyper-volatile ice content. Unlike previous predictions, we show that hyper-volatile survival was possible in a significant number of planetesimals of the Kuiper Belt and the Oort Cloud. Planetesimals ejected into the interstellar space proved to be the most processed,  
while planetesimals ending in the Oort Cloud were the least processed population. We show that processing differences between populations are a direct consequence of their orbital evolution patterns, and that they provide a natural explanation for the observed variability in the abundance ratios of CO on cometary populations and on the recent observations of long-distance CO-driven activity on inbound Long-period Comets.

\end{abstract}



\section{Introduction} \label{sec:Intro}

Comets have captivated the interest of the scientific community for decades due to their presumed pristine nature and their ability to provide valuable information on the conditions of the primordial Solar System and its subsequent evolution \citep[e.g.][]{Prialnik_1995, De_Sanctis_2015, Bockelee_2015, Wierzchos_2020}.

On a crude outline of a comet's lifetime, we can identify three main evolutionary phases \citep{Meech_2004}.  First, future cometary nuclei formed as ice-rich planetesimals in the outer parts of the Sun's gas-dominated protoplanetary disk~\citep{simon2022}.  Their orbits evolved under the effects of gas drag~\citep{Adachi_1976} and of the growing planets.  The giant planets' dynamical instability destabilized the orbits of cometary nuclei, lead them to be strongly gravitationally scattered and, in large part, ejected from the Solar System.  A fraction of `lucky' planetesimals were trapped in two outer Solar System reservoirs: the Kuiper Belt and its extension the scattered disk~\citep[hereafter KB/SD;][]{Levison_2008}, and the Oort Cloud (hereafter OC) \citep[e.g.][]{Dones_2015, Nesvorny_2018, Kaib_Volk_2022}. The second phase of evolution of comets involves a prolonged residence in the reservoirs, on relatively stable orbits for timescales of the order of billion years. Due to the prevailing low temperatures in these far-away reservoirs, planetesimals are generally considered to have remained unaltered, conserving their primordial physical and chemical characteristics, and justifying their characterization as the most primitive objects of our Solar System \citep{Weissman_2020}. In the third phase, these planetesimals return to the inner Solar System\footnote{Here the terms `inner' and `outer' describe the areas within and beyond the orbit of Neptune respectively.}, and can then be called `comets'. During this phase, comets are usually detected and observed as they undergo extensive thermal processing due to their gradual approach to the Sun.  This final phase almost universally culminates with ejection into interstellar space.

The view that comets are `pristine' and largely unaltered has come into question. A number of recent studies have suggested that thermal processing, and the resulting activity and alterations, may have started during the early phases of comets' evolution, during their prolonged reservoir residence \citep[e.g][]{Haruyama_1993, Prialnik_1995, De_Sanctis_2001, Choi_2002, Kral_2021, Parhi_2023}, or before reaching the inner Solar System, in the giant planet area \citep{Guilbert-Lepoutre_2012, Guilbert-Lepoutre_2014, Jewitt_2021, Gkotsinas_2022, Kaib_2022}. This claim is supported by numerous observations of long-distance activity on Centaurs \citep[e.g][]{Jewitt_2009}, or inbound Long-period Comets \citep[e.g][]{Meech_2017, Hui_2019, Farnham_2021}. 
Some studies \citep{Raymond_2020, Davidsson_2021, Lisse_2022} have even suggested processing at the very early stages of comets' lifetimes, within the planetesimal disk and during their implantation into cometary reservoirs. All these studies, even if they do not question the primitive status of comets, point to the necessity of reassessing it, in order to better interpret current observations and improve theoretical models and predictions. 

In this paper, we re-evaluate the primitive nature of comets. We propose a theoretical framework to model the early thermal processing of ice-rich planetesimals that eventually either are ejected from the Solar System or return as Jupiter-family Comets (hereafter JFCs) or Long-period Comets (hereafter LPCs), during their implantation into the \kb and the \ooc. Our framework consists of an adapted thermal and ice evolution model, coupled to orbital trajectories of implantation or ejection obtained from \textit{N}-body simulations. These allow us to couple the effects of thermal evolution with the comets' orbital evolution, an aspect that has been neglected in most previous studies. By using a large number of particles, we provide a statistically significant first-order assessment of the processing sustained to the three planetesimal populations (\kb, \ooc and ejected planetesimals) during the first phase of their lifetime, allowing us to make predictions regarding their future evolution in the reservoirs and on their return to the inner Solar System. 

In Section \ref{Sec: N-body simulations} we provide a brief description of the \textit{N}-body simulations and the planetesimal samples used in this work. In Section \ref{sec:Methods}, we detail the adapted thermal and ice evolution models developed for this study and we explain how the coupling between these models and the \textit{N}-body simulations is achieved. 
In Section \ref{sec:Results} we present the results of our simulations and in Section \ref{Sec:Discusssion} we discuss their implications on the processing of the three studied populations, while analyzing the limitations of our approach and their consequences. A brief summary of the main points of this work is provided in Section \ref{Sec: Conclusions}.


\section{\textit{N}-body simulations} \label{Sec: N-body simulations}

We used orbital trajectories obtained from two different sets of \textit{N}-body simulations. Both sets simulated the orbital evolution of planetesimals during the giant planet instability~\citep{Nesvorny_2018}, and also included torques from the Galactic tide~\citep{Heisler_1986} in order to enable the capture of planetesimals in the Oort cloud. Both sets of simulations invoked the existence of an additional ice giant planet that was ejected during the instability~\citep{Nesvorny_2011, Nesvorny_2012}, and both put stringent limits on the simulations that were deemed successful in terms of their matches to the present-day planets' orbits~\citep[see, for example, ][]{Clement_2021}. The first set of simulations, described in detail in \citet{Nesvorny_2017} and \citet{Vokrouhlicky_2019}, provided two samples of planetesimals that eventually populated the \kb, and the \ooc. The second set of simulations, detailed in \citet{Raymond_2020}, provided a sample of planetesimals that were ejected of the Solar System. In the following paragraphs, we briefly describe the main considerations and initial conditions of these simulations, and give a detailed description of the samples. We refer the reader to the aforementioned publications for a more comprehensive review. 

Both sets of \textit{N}-body simulations tracked the orbital evolution of the four giant planets, placed initially in a multi-resonant setup \citep{Morbidelli_2007}, and of an additional ice giant, introduced to increase the probability of reproducing the Solar System's current orbital architecture \citep{Nesvorny_2012}. A number of massless particles (10$^6$ in the first and 10$^3$ in the second simulation) was placed on low-eccentricity ($<$0.1) and low-inclination ($<$0.05) orbits, on an outer disk, extending from a heliocentric distance of 24~au to 30~au in the first set of simulations, and from 21.4~au to 30~au in the second set of simulations. The initial internal time step was set at 0.5 yr for the first simulation and at 24 days for the second one. The influence of terrestrial planets was not considered in any simulation, nor were the effects of non-gravitational forces. 

The simulations started from the moment the giant planets' instability was triggered. The first was integrated over the age of the Solar System, until the present epoch, although for this study only a fraction of it (200~Myr) was considered, as it proved sufficient for the formation of the \kb and the \ooc reservoirs. A modified version of the \texttt{swift\_rmvs4} code, part of the Swift \textit{N}-body integration package \citep{Levison_1994}, was used to track the orbital evolution of planets and planetesimals. In addition to the nominal output frequency of the Swift integrator (1 Myr), the orbital elements of all planetesimals were recorded with a cadence of 1000~yr, increased to 100~yr whenever a planetesimal was detected at heliocentric distances within 23~au, where increased thermal processing was anticipated. 

The second set of simulations was integrated for 1~Gyr after the instability, using a modified version of the \textit{Mercury} hybrid integrator \citep{Chambers_1999}. The nominal output frequency for the orbital elements of planetesimals was set at 1~Myr. An additional work to increase this cadence was done, using a track record of close gravitational encounters between the planetesimals and the giant planets, and a track record of passages at distances within 2.5~au of the Sun. To do so, we assumed that the semimajor axis of the planetesimals' orbit remained stable between close encounters \citep{Milani_1987}, and that any orbital changes were due to small modifications in the eccentricity. This allowed us to reconstruct individual pathways leading to ejection with a denser, yet non-constant, output frequency that provided a better time resolution.

\subsection{The Kuiper Belt - Scattered Disk Sample}

The \kb sample consists of 383 planetesimals. They were selected at the end of the simulation, which corresponds to the current epoch, by comparing the simulated orbital elements to the actual orbital distribution of observed JFCs, obtained from the JPL Small-Body Database Search Engine\footnote{\url{ http://ssd.jpl.nasa.gov/sbdb_query.cgi}}. This choice ensured that the selected planetesimals would eventually return to the inner Solar System as ``active'' JFCs, with perihelia within 2.5~au, at least once in their lifetime, and a known total absolute magnitude ($H_t$) smaller than 10.9, two conditions that allowed a better characterization of the sample.

\begin{figure}
    \centering
    \includegraphics[scale=0.8]{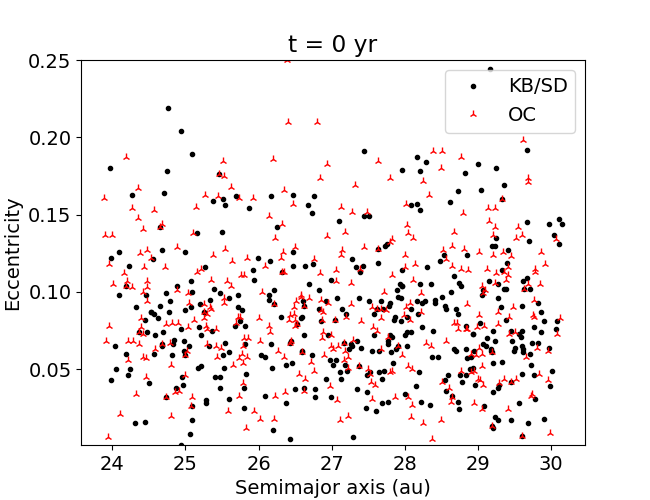}
    \includegraphics[scale=0.8]{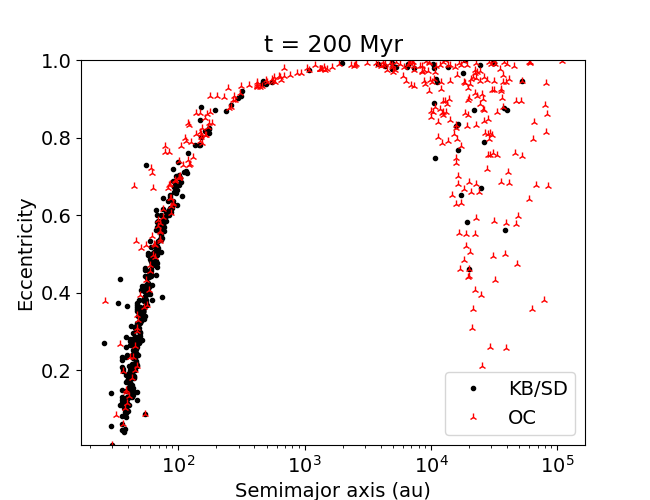}
    \caption{Orbital distribution in the semimajor axis and eccentricity plane of the \kb (black dots) and \ooc (red triangles) planetesimals at the beginning (t=0 yr) (top panel) and the end of the simulation (t=200 Myr) (bottom panel).}
    \label{fig:KB_OC_distribution}
\end{figure}

Figure \ref{fig:KB_OC_distribution} shows the distribution of the \kb sample on the semimajor axis-eccentricity plane at the beginning (t=0~yr, i.e. the timing of the giant planet instability) and what we considered to be the end of the simulation in this work (t=200 Myr). We note that the largest fraction of planetesimals ($\sim$47\%) populates the inner Scattered Disk, defined by orbits with 50 $<a<$ 200~au. A fraction of 41\% of planetesimals ends up in the classical Kuiper Belt ($a<$ 50~au), and a small percentage (2\%) ends up at the outer scattered disk (200 $<a<$ 1000~au). 
Interestingly, a non-negligible number of planetesimals ($\sim$9\%) ends up in the Oort Cloud  ($a>$ 1000 au) on low-inclination orbits ($<$1.0\textdegree), from which they are able to return into short-period and low-inclination orbits \citep{Emelyanenko_2013, Nesvorny_2017}.

In the upper panel of Figure \ref{fig:OC_orbital_evolution}, a typical example of a planetesimal's implantation in the inner scattered disk is presented. Departing from a low-eccentricity ($e\simeq0.1$) orbit with a semimajor axis of $\sim$28~au, this planetesimal evolved dynamically under the influence of Neptune. Through a series of close encounters its orbital elements were excited until the planetesimal reached the inner parts of the scattered disk ($a\simeq$ 90-100 au), where it settled to high-eccentricity orbits ($e\simeq$ 0.5-0.7) after the first $\sim$25~Myr of the simulation.

\subsection{Oort Cloud Sample}

The \ooc sample consists of 360 planetesimals. It was calibrated with a sample of LPCs so that:
(1) the sample was of comparable size to the \kb sample, and (2) upon their return to the inner Solar System, all its members had minimum perihelion distances within 10~au, with a linear cumulative distribution function. These arbitrary conditions were chosen due to the lack of a comprehensive survey for the orbital distribution of LPCs \citep{Vokrouhlicky_2019}.
Figure \ref{fig:KB_OC_distribution} shows the initial and final distribution of the \ooc sample as well. Despite the relatively small number of planetesimals, we can observe the formation of the Oort Cloud after 200~Myr. At that moment, the majority of particles (61\%) is already placed in the Oort Cloud ($a>$1000~au), populating for the most part (75\%) the region beyond the Oort peak at 10,000~au. Of the remaining 39\%, some planetesimals are observed to be on transient orbits with $a>$ 600~au and high eccentricities, while some populate the region with $a<$ 1000~au, from where they are expected to return to the inner solar system as Halley-type comets.

\begin{figure}
    \centering
    \includegraphics[scale=0.8]{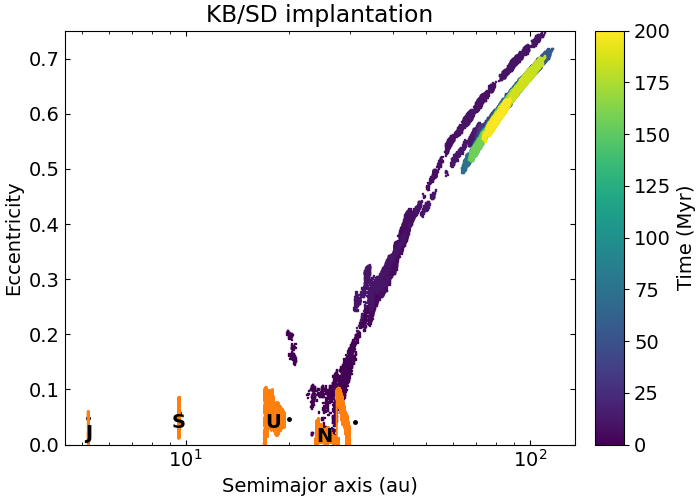}
    \includegraphics[scale=0.8]{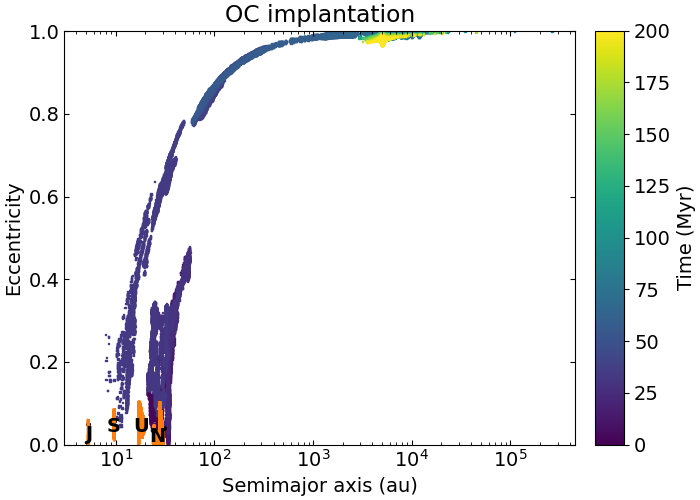}
    \caption{Example of planetesimals' implantation on the Kuiper Belt (upper panel) and the Oort Cloud (lower panel) on the semimajor axis vs eccentricity plane. The color code provides the timescale of the simulation. The orange clusters denote the orbital evolution of Jupiter, Saturn, Uranus and Neptune.}
    \label{fig:OC_orbital_evolution}
\end{figure}

The lower panel of Figure \ref{fig:OC_orbital_evolution} shows a typical example of orbital evolution for a planetesimal of this sample as it is implanted in the Oort Cloud. It started on a low eccentricity ($e\simeq0.08$) orbit with a semimajor axis of $\sim$28~au and then, after a series of encounters with Neptune and Uranus, this planetesimal was scattered inward, and came under the dynamical control of Saturn after $\sim$50~Myr. Through a series of close encounters with Saturn, it was eventually scattered toward the Oort Cloud rather quickly, $\sim$60~Myr after the start of the simulation. It reached an orbit beyond the Oort peak before it settled in the inner Oort Cloud ($a<$10,000~au) on a high-eccentricity orbits.

\subsection{Ejected Planetesimals} \label{SubSec:Ejected Pop}

The sample of ejected planetesimals consists of 675 particles. These planetesimals interacted with the giant planets after the disk was dispersed and were eventually ejected out of the Solar System without ever reaching one of the reservoirs. A planetesimal was considered to be ejected when it reached distances larger than 1~pc (206,265~au). Although this simulation was integrated for 1~Gyr, the ejection timescales were in general significantly shorter and varied between particles, as each planetesimal's trajectory is unique and determined by the stochastic nature of its encounters with the giant planets. On average, a planetesimal was ejected after $\sim$32~Myr (the median time was $\sim$17 Myr) from the beginning of the simulation. The first recorded ejection took place at $\sim$1~Myr, and the last at $\sim$186~Myr.


\section{Dedicated thermal and ice evolution models} \label{sec:Methods}

Our goal was to evaluate the degree of thermal processing of planetesimals during the phase of implantation or their ejection from the Solar System. To this end, a dedicated thermal evolution model was designed, adapted to the needs of the coupling to the trajectories proposed by the \textit{N}-body simulations. This thermal evolution model, for reasons that will become obvious in the current Section, focuses mainly on two main processes of the thermally induced evolution: the energy and gas diffusion inside a porous nucleus. In this Section we describe the key elements and assumptions of our model and the numerical limitations that led us to the latter. 

\subsection{Thermal evolution model} \label{Subsection:Thermal evolution model}

To describe the heat diffusion inside the nucleus of the simulated planetesimals, our thermal evolution model solves the energy conservation equation:
\begin{equation}
\rho_{bulk}c~ \frac{\partial T}{\partial t} + div(-\kappa ~\overrightarrow{grad}~ T) = \mathcal{S} \label{Equation:Energy conservation equation}
\end{equation}
\noindent where $\rho_{bulk}$ (kg~m$^{-3}$) is the nucleus' bulk density , $c$ (J~kg$^{-1}$~K$^{-1}$) the nucleus' material heat capacity, $T$ (K) the temperature, $\kappa$ (W~K$^{-1}$~m$^{-1}$) the nucleus' effective thermal conductivity. The term $\mathcal{S}$ corresponds to the collective contributions of internal heat sources and sinks, related to a variety of physical processes such as the latent heat consumed or released during the sublimation and condensation of ices \citep[e.g.][]{Benkhoff_1991}, heat transport by gas diffusion in the porous matrix \citep[e.g.][]{Benkhoff_1995}, internal heating by radioactive elements \citep[e.g.][]{Prialnik_1987} or energy released during the exothermic process of crystallization of amorphous water ice \citep[e.g.][]{Schmitt_1989}. 

The planetesimals were modeled as spherical, porous aggregates of dust and ices, with a radius of 5~km. The porosity was fixed to a constant value of 75\%, based on porosity estimates for 67P/Churyumov-Gerasimenko \citep{Sierks_2015, Jorda_2016}. Thermophysical parameters were chosen as typical average values from the published literature \citep[e.g.][]{Huebner_2006}. For the dust component a mixture of refractory and organic materials was considered. For the icy component, a selection of the more abundant ice species detected on comets \citep[e.g.][]{Bockelee_2017} was selected: CO -- which was also considered representative of hyper-volatile species, \dioxi -- representative of moderate volatile species, and \water ice in -at least initially- amorphous form. Given the broad range of values and uncertainties regarding the abundances of those species on cometary populations \citep[e.g.][]{Ootsubo_2012, Pinto_2022} or individual comets \citep[e.g.][]{Choukroun_2020, Biver_2021}, a unique, simplified initial composition was chosen. The refractories-to-ice mass ratio was set to one, a slightly lower value than the predictions for ice-rich planetesimals at the beginning of the simulations \citep[e.g][]{Choukroun_2020}. The relative abundances to \water in the model were preset as follows: CO/\water = 5\% and CO$_2$/\water = 15\%, implying an initial composition of 42\% \water ice, 6\% \dioxi ice and 2\% CO ice.

The terms appearing on the left-hand side of Equation \ref{Equation:Energy conservation equation} ($\rho_{bulk}$, $c$ and the conductivity $\kappa$ of the solid component) were modeled as mass-weighted averages of the individual values of their respective components as those were presented in the previous paragraph. A selection from the most up to date values available in the published literature was used wherever possible. This selection is summarized in Table \ref{Table:Physical Parameters} of the Appendix. Appropriate corrections to account for the porous nature of the nucleus were introduced following typical considerations for the modeling of cometary nuclei \citep{Prialnik_2004, Huebner_2006}.

Of these terms, the most crucial in our thermal evolution model was the effective thermal conductivity, which controlled the amount of energy transferred toward a planetesimal's interior. A variety of theoretical approximations for the calculation of the effective thermal conductivity exists in the published literature. For a comprehensive review of the different approximations and their historical evolution we refer the reader to \cite{Prialnik_2004} and \cite{Huebner_2006}. In this work, we used Russel's formula \citep{Russell_1935} as a first reduction factor $\phi$ to account for the porous nature of cometary material (see also Appendix \ref{Sec:Appendix}). Considering that the solid matrix is composed of small icy dust grains, an additional reduction known as the Hertz factor $h$ was introduced to describe conduction only through a small fraction of the grains' total surface area. This correction is calculated as the ratio of the contact area to the cross section of a spherical grain \citep{Kossacki_1999}. Laboratory measurements give a wide variety of values, ranging from 10$^{-1}$ to 10$^{-4}$ \citep{Huebner_2006}. However a consensus seems to have been established around values of the order of 10$^{-2}$ \citep[e.g.][]{Huebner_2006, Gundlach_2012}, which is the value used in the model.

Accounting for the above corrections and the contribution of the radiative transport through the pores given in Equation (\ref{Equation:Conductivity_Radiation}), the final expression for the effective thermal conductivity is: 
\begin{equation}
   \kappa_{eff} = h \phi \kappa_{solid} + \kappa_{rad} \label{Equation:Effective Thermal Conductivity}
\end{equation} 
The conductive term in Equation (\ref{Equation:Effective Thermal Conductivity}) dominates over the radiative term at low temperatures (T$<$100~K). Above this threshold the contribution of the radiative term becomes more significant, and dominates above 150~K, as the heat is mainly transported by radiation through the pores \citep{Gkotsinas_2022}. For a porosity of 75\% and a Hertz factor $h$ = 10$^{-2}$ our resulting thermal conductivity falls in the 10$^{-3}$--10$^{-4}$~W~K$^{-1}$~m$^{-1}$ range, in good agreement with estimates for comet 67P/Churyumov-Gerasimenko \citep{Keller_2015}.

\subsection{Coupling thermal calculations to \textit{N}-body simulations} \label{SubSec: Coupling}

The coupling between a planetesimal's orbital and thermal evolution is made through the energy balance at the surface, i.e. the surface boundary condition of the thermal evolution model:

\begin{equation}
    (1-\mathcal{A}) \frac{L_\odot}{4\pi d_H^2} \cos \zeta = \varepsilon \sigma T^4 + \kappa \frac{\partial T}{\partial z} \label{boundary_eq}
\end{equation}

\noindent where the insolation on the left-hand side of the equation is a function of the Bond's albedo ($\mathcal{A}$), the solar constant ($L_\odot$), the heliocentric distance ($d_H$ in au) and the local zenith angle ($\zeta$). We note that the solar constant is reduced to a percentage of its current value (0.8$L_\odot$), where $L_\odot = 1360 W m^{-2}$, in a coarse attempt to account in a simplified manner the changes of the solar luminosity in the early stages of our Solar System \citep{Gough_1981, Lagarde_2012}. More elaborated approximations exist and have been employed on other models \citep[see for example][]{Davidsson_2021}, but we consider that for a first-order assessment of the early thermal evolution this simplified value is sufficient. Moreover, the local zenith angle is set to zero, meaning that our simulations focus on the processing around the sub-solar point. This choice implies that we are focusing on the maximum effect of the solar illumination, as it is quantified for the purpose of this work (see \ref{SubSec: Coupling}). This end-member scenario is not necessarily characteristic of other areas of cometary nuclei, such as the poles, where lower temperatures, and as consequence less processing, are expected. The thermal emission, on the right-hand side of the equation, is a function of the emissivity ($\varepsilon$), the Stefan-Bolzmann constant ($\sigma$) and the temperature (T in K), and the heat flux toward the interior of the thermal conductivity ($\kappa$ in W~m$^{-1}$~K$^{-1}$) at the surface. 

More specifically, the coupling is realized through an accurate calculation of the heliocentric distance ($d_H$) at each and every time step of the dynamical simultions, appearing in the calculation of the received solar energy. Obviously, such a calculation is not straightforward, as information on the position of planetesimals is only available at specific intervals.

In long-term simulations, it is a common practice to replace eccentric orbits with circular ones, assuring the conservation of a key quantity, such as the total amount of solar energy received over one orbital period \citep[e.g.][]{Prialnik_2009, Guilbert-Lepoutre_2012, Gkotsinas_2022, Lilly_2024}.
Although such an approximation results in loss of information, mainly related to perihelion passages and seasonal activity, it provides not only a non-negligible improvement in numerical efficiency, but also a very elegant solution to a variety of issues raised by the use of eccentric orbits. For example, it lifts the ambiguity introduced by the lack of information on the direction of the comet nucleus, i.e. inbound vs outbound, during a given calculation timestep \citep{Safrit_2021}.
In this work, following a previous study \citep{Gkotsinas_2023}, we employ the effective thermal distance $r_T$, an expression proposed by \cite{Mendez_2017}, following previous similar calculations for the annual mean insolation and the effective surface temperature \citep[e.g.][]{Ward_1974}, producing circular orbits with the same average equilibrium temperature for an eccentric orbit with a given semimajor axis ($a$) and eccentricity ($e$):

\begin{eqnarray}
    r_T &=& a \left[ \frac{2 \sqrt{1+e}}{\pi} \textit{\textbf{E}} \sqrt{\frac{2e}{1+e}} \right]^{-2}\\
        &\approx& a (1 + \frac{1}{8}e^2 + \frac{21}{512}e^4 +\mathcal{O}(e^6) ) \label{Equation:Equilibrium_temperature}
\end{eqnarray}

\noindent with \textit{E}; the complete elliptic integral of the second order. 
As explained, this approximation does not allow the resolution of seasonal variations of the insolation related to eccentricity, nor the consideration of seasonal variations related to obliquity. The same goes for diurnal variations, which are unique to each planetesimal and cannot be studied in a statistically significant manner.

\subsection{Internal energy sources and sinks} \label{SubSec:Sources and Sinks}

\begin{figure}
    \centering
    \includegraphics[width=\linewidth]{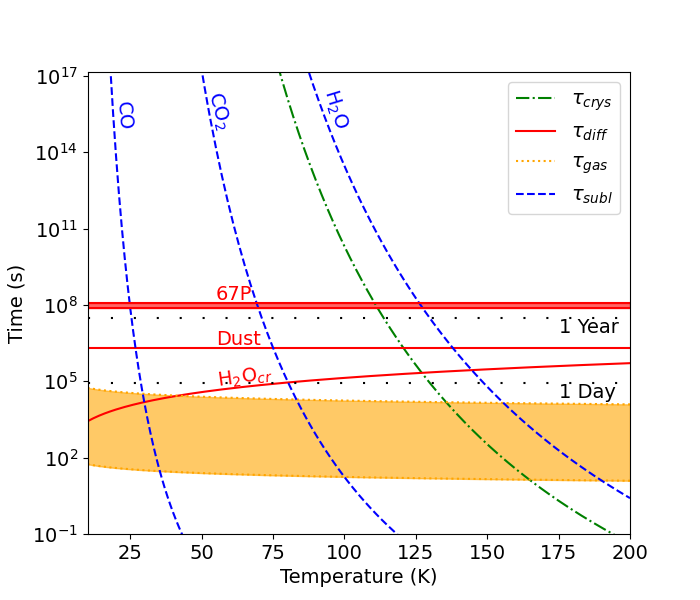}
    \caption{Characteristic timescales of the main physical processes taking place in a comet's nucleus: (a) Crystallization of amorphous \water ice denoted by the green dashed dotted line, (b) Sublimation of different volatile species denoted by the blue dashed lines, (c) Gas diffusion within the pores for pores' radius between 10$^{-3}$ and 10$^{-6}$ denoted by the orange dotted lines and the orange shaded area, (d) Heat diffusion in nucleus of different compositions: pure dust, pure crystalline \water ice and a nucleus with characteristics resembling those of comet 67P/C-G. The black loosely dotted lines serve as timescale indicators. Figure adapted from \cite{Prialnik_2004, Huebner_2006}.}
\label{fig:Characteristic timescales}
\end{figure}

While the orbital elements in the \textit{N}-body simulations are recorded with a relatively large cadence (e.g. 100 to 1000~yr in the first set of simulations), thermal processes usually operate on much shorter timescales. 

In Figure \ref{fig:Characteristic timescales}, these timescales are plotted as a function of temperature, for a hypothetical subsurface layer of thickness $\Delta r$=1~m, with a porosity of 50~\%, typical cometary values for the bulk density, the heat capacity and the effective thermal conductivity, and for different compositions. We can easily observe the wide range of timescales on which each process operates, accelerating significantly as the temperature rises, and reducing the timescale by several orders of magnitude. In addition, we observe important differences when these processes are compared to each other at a given temperature. A striking example is the comparison between the thermal and the gas diffusion timescales, the former varying between a few months (10$^6$~s) and a little over year depending on the considered composition, while the latter takes place in a matter of hours or minutes depending on the size of pores. Sublimation is an exponential function of the temperature: \dioxi and \water ices sublimate in a few minutes at $\sim$80~K and $\sim$180~K respectively, but are practically insignificant below these threshold temperatures. The time needed for the crystallization of amorphous water ice at temperatures below $\sim$80 K, exceeds the Solar System's lifetime, but can take place in just a few years for temperatures between 100 and 110~K, and in only a few hours for temperatures between 125-150 K \citep{Jewitt_2009, Prialnik_2022}. Accounting for processes with such diverse timescales has already been one of the main challenges of thermal evolution models \citep[e.g.][]{Orosei_1999, Prialnik_2009}, even when the integration period was short and only a single object was studied. In this work, as we dealt with long-term simulations, requiring substantial time of integration and as we used a large number of objects to get statistically significant image of the processing, further simplifying assumptions were required. 

We fixed the time step for thermal calculations at one year, unless smaller values were available in the case of the ejected planetesimals. This means that the thermal evolution model created smaller time steps within each \textit{N}-body simulations' output, that allowed a more accurate calculation of the internal temperature of the planetesimals. This value was considered a good compromise between the different timescales involved in this study. It allowed the resolution of the heat diffusion in the interior of the planetesimals, a process operating on timescales of a little over a year for a planetesimal with properties close to those of comet 67P (thick red line in Figure \ref{fig:Characteristic timescales}), while being comparable to the \textit{N}-body simulations' output cadence. On the downside, the other thermal processes could not be resolved and they were not taken into account in our thermal evolution model. The possibility of heating due to the decay of radioactive elements was not considered as well, as on the one hand the presence of radioactive elements remains uncertain \citep[e.g.][]{Levasseur_2018, Malamud_2022} and on the other hand as the timing of the instability, and therefore the inclusion of radioactive elements on planetesimals, is still a matter of debate \citep[e.g.][]{Raymond_2020, Golabek_2021}. 

To summarize, the current configuration of our thermal evolution model neglects phase transitions, at least in a classic direct way, implying that the term $\mathcal{S}$ on the right-hand side of Equation \ref{Equation:Energy conservation equation} is zero. As a consequence, it does not describe faithfully periods of cometary activity, when thermally-induced phase transitions contribute significantly to the total energy budget. Ideally, this should be obtained by the simultaneous resolution of a mass conservation equation \citep[e.g.][]{Davidsson_2021}, the results of which are used in the sources and sinks term of the energy conservation equation (Equation \ref{Equation:Energy conservation equation}). 
Given the long time steps of our simulations, this classic resolution scheme was not evident, so we opted for an alternative approach to get an estimation of the ice content evolution. This approach, consisted in the asynchronous coupling of an ice evolution model to the thermal evolution model, a solution initially proposed by \cite{Schorghofer_2008} and further developed in \cite{Schorghofer_2010, Schorghofer_2016} and \cite{Schorghofer_2018}.

\subsection{Asynchronous ice evolution model} \label{SubSec:Asynchronous Model}

The basic idea behind the ice evolution model is that we were dealing with temperature-driven processes. Therefore, by calculating an average of the temperature variations over a given time interval (for example a time step of 100 or 1000~yr) can give us an estimation of the evolution of the ice content over this interval, by calculating the ensuing average vapor pressures developing in a planetesimal's interior. In the following paragraphs, a brief overview of the ice evolution model is provided. We refer the reader to the aforementioned publications for a more detailed description.

Assuming an ice interface buried at some depth at a radial distance from the center of a planetesimal's nucleus ($r_i$), following \cite{Schorghofer_2018}, we write the vapor flux at the interface as:
\begin{equation}
   J(r_i) = \rho_i \frac{dr_i}{dt} \label{Equation:Vapor Flux}
\end{equation}
where $\rho_i$ (kg~m$^{-3}$) is the ice density. 

Supposing that the vapor flux is proportional to $1/r^2$, we write the vapor flux at a random radial distance from the center as:

\begin{equation}
    J(r) = -D \frac{d \rho_v(r)}{dt} = \left(\frac{r_i}{r} \right)^2 J(r_i) \label{Equation:Vapor_at_r}
\end{equation}
where $D$ (m$^2$s$^{-1}$) is the vapor diffusion coefficient and $\rho_v$ (kg~m$^{-3}$) the vapor density. The vapor diffusion coefficient can be expressed as:
\begin{equation}
    D = \frac{\pi}{8 + \pi} \frac{\psi}{1 - \psi} \frac{v_{th}}{\xi} r_p \label{Equation:Local Diffusion Coef}
\end{equation}
with $\psi$ the nucleus' porosity, $v_{th}=\sqrt{8 k_b T / \pi m}$ (m s$^{-1}$) the mean thermal speed, $r_p$ (m) the pores' radius and $\xi$ the tortuosity, an intrinsic property of the porous medium defined as the ratio between the actual flow path to the straight distance between the two ends of this flow path \citep{Bear_1988}. This term seeks to describe the `twisting' of the pores in a porous medium structure and here is set to one, implying that the pores, or the flow path, is considered to be a straight line, a very common assumption in thermal evolution models \citep{Huebner_2006}. 
It should be noted that other expressions for the vapor diffusion coefficient exist, such as the Knudsen's formula \citep{Espinasse_1991} or the Clausing equation \citep{Steiner_1990}, but as \cite{Guilbert-Lepoutre_2014} reported, the calculated vapor fluxes differ by a factor that does not exceed the order of unity.

Resolving Equation \ref{Equation:Vapor_at_r} for the vapor density, and subsequently deriving it at $r =r_i$, supposing that the vapor pressure at the surface is zero and at the interface equal to the saturation vapor density ($\rho_s$), we obtain the following expression for the vapor flux at the ice interface \citep{Schorghofer_2018}:

\begin{equation}
    J(r_i) = \frac{D \rho_s}{r_i(1 - r_i/R)} \label{Equation:Vapor flux 2}
\end{equation}

Combining Equation \ref{Equation:Vapor Flux} and \ref{Equation:Vapor flux 2}, we can obtain an expression for the retreat rate of the ice interface, in the form of a differential equation that can be integrated to a time-stepping scheme:

\begin{equation}
    \frac{dr_i}{dt} = D \frac{\rho_s}{\rho_i} \frac{1}{r_i(1 - r_i/R)}
\end{equation}
or more conveniently:

\begin{equation}
    r_i \left(1 - \frac{r_i}{R} \right) dr_i = D \frac{\rho_s}{\rho_i}dt \label{Eqution:Retreat_rate}
\end{equation}
we note that the saturation vapor pressure ($\rho_s$), depends only on the temperature and can be calculated from the saturation vapor pressure ($P_{sat}$ (Pa)) using the average temperatures provided by the thermal evolution model at each time step provided by the dynamical simulations (100 or 1000 yr for \kb and \ooc objects and 1 Myr or smaller for ejected planetesimals).

Here, we calculate the saturation vapor pressure using a simple empirical expression obtained from the Clausius-Clapeyron equation for the \dioxi and the \water ice molecules \citep{Fanale_1984, Fanale_1990}: 
\begin{equation}
    P_{sat}(T) = A e^{-B/T} \label{Equation:Vapor pressure}
\end{equation}

\noindent where the coefficients $A$ and $B$ are 107.9$\times$10$^{10}$ Pa and 3148 K for \dioxi, and 356$\times$10$^{10}$ Pa and 6141.667 K for \water respectively. This expression, although relatively simple, is in excellent agreement with more complex formulas as the ones proposed by \cite{Brown_1980} and \cite{Orosei_1999}. However, for CO, this expression, using coefficients from the same work \citep{Fanale_1984}, presents important discrepancies with more complex ones \citep{Brown_1980, Fray_2009} and with the results of recent laboratories measurements \citep{Grundy_2023}, that measured significantly lower values, by at least two orders of magnitude, for temperatures below 30~K. In this study, an expression proposed by \cite{Brown_1980} is employed, as it is considered a good compromise between the more complex expressions and the aforementioned laboratory measurements:

\begin{equation} \label{Eq:Saturation Vapor CO}
ln(P_{sat}) = A_0 + \sum_{i=1}^5 A_i T^i
\end{equation}

\noindent where the values for the coefficients $A_0$ and $A_i$ are given in Table \ref{Table:Saturation Pressure Coeff} of the Appendix. 

Obviously, this analytical method does not explicitly account for internal processes such as the gas flow or the recondensation below and above the ice interfaces. It does however remain a powerful tool for a first order estimation of a planetesimal's evolution as it still takes into account the two dominant processes of the thermally induced activity: the heat and the gas diffusion inside the nucleus.

\subsection{Supplementary ending conditions} \label{SubSec:Ending Conditions}

As we mentioned earlier, for this work, the orbital evolution of the planetesimals was recorded for a period of 200 Myr or until the ejection of a planetesimal from the Solar System. For the thermal processing simulations, we added three supplementary termination conditions to further optimize the numerical efficiency. 

The first condition was introduced to terminate a simulation when no thermal modification was recorded after a number of iterations. A threshold of 100 iterations was defined, after which if no modification of the ice interfaces was recorded, the simulation ended. This condition targeted planetesimals that quickly consumed their hyper-volatile content, or that were rapidly placed in far-away, relatively stable orbits, where the temperatures were too low for any further activity to take place. 

The second condition forced the termination of the thermal simulations if a planetesimal's semimajor axis exceeded 100 au. This condition concerned planetesimals that managed to ``maintain'' some minor activity, mainly due to small changes in the float precision, ``escaping'' in a way the first condition. As these changes concern mostly planetesimals orbiting at large heliocentric distances ($>$70~au), the 100~au limit was considered a safe choice. An estimation of a black body's temperature at 100~au, for a a luminosity value of 0.8$L_\odot$ yields $\simeq$22 K, well below the sublimation temperatures for \dioxi and \water, and at the limit for CO which however was never present close to the surface at these distances. We verified that no further orbital changes reducing significantly the semimajor axis of the planetesimals occurred after the termination of our thermal simulations. Extending this limit to 200 au did not provoke any substantial change in the outcome of our simulations.   

Finally, a termination condition targeting the ejected planetesimals was added, ending the thermal evolution simulations when their orbits became hyperbolic, flagged by eccentricity values above one. This choice was made first and foremost for consistency with our orbital averaging technique, valid only for bound, eccentric orbits. Either way, from a thermal evolution point of view, it was a ``harmless'' condition since the transition from eccentric to hyperbolic orbits generally occurs at large heliocentric distances where no significant thermal alteration was recorded.

The application of these ending conditions, as expected, reduced the time of the thermal evolution simulations: the evolution of \kb planetesimals was simulated for $\sim$63 Myr on average (median time of $\sim$20 Myr), with the shortest simulation lasting $\sim$1.4 Myr. For the \ooc sample, the corresponding numbers were considerably lower, $\sim$23 Myr and $\sim$11 Myr for the mean and the median simulation timescales respectively, and $\sim$640~kyr for the shortest simulation. These numbers provide a first hint on the differences between the two populations: planetesimals scattered toward the Oort Cloud were not only scattered further out, but also much faster. The effect was similar on the ejected planetesimals: a significant decrease on the thermal simulation timescale. On average a planetesimal reached a hyperbolic orbit in $\sim$15~Myr, with the median value reduced to $\sim$10~Myr. The shortest ejection was recorded to take place in just $\sim$1000~yr.

\section{Results} \label{sec:Results}

\subsection{Individual planetesimal example}

\begin{figure*}[ht!]
    \centering
    \includegraphics[width=\textwidth]{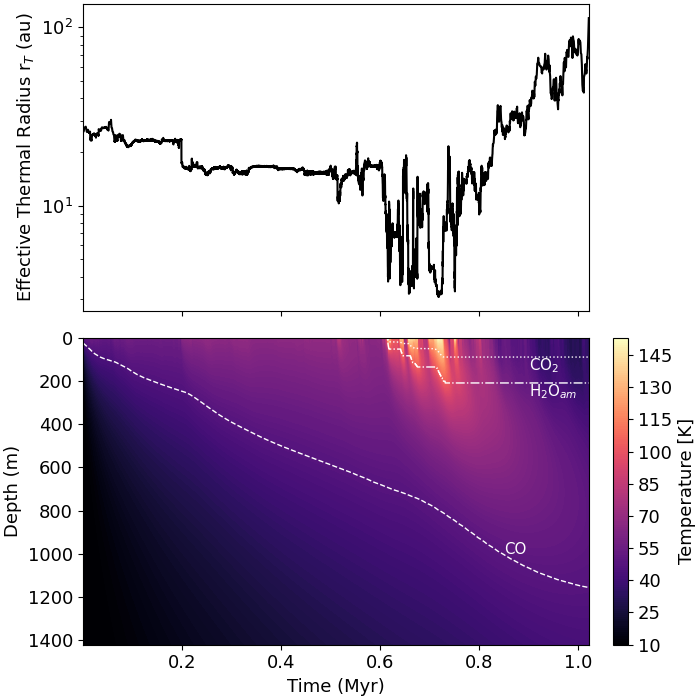}
    \caption{Thermal processing of a planetesimal scattered toward the Oort Cloud. Upper panel: Evolution of the effective thermal radius (r\textsubscript{T}). Lower panel: Temperature distribution (denoted by the color code) in the planetesimal's interior at a subsurface layer of $\sim$1000~m. The dashed, dotted and dashed-dotted white lines show the CO, \dioxi and crystallization interfaces respectively.}
\label{fig:Processing_LPC}
\end{figure*}

Figure \ref{fig:Processing_LPC} shows an example of the coupling between the thermal and ice evolution models and the orbital evolution of a planetesimal scattered toward the Oort Cloud. We plot data relative to a 1.4 km subsurface layer only, allowing a better visualization of the temperature and ice interfaces' evolution. The thermal evolution of this planetesimal was computed for a period of $\sim$1 Myr, before the planetesimal was scattered beyond the threshold of 100 au. 
The lower panel of Figure \ref{fig:Processing_LPC} displays the temperature distribution in the planetesimal's interior, calculated by the thermal evolution model. Plotted over this temperature distribution are the evolution of the positions of the three ice interfaces, as calculated by the ice evolution model. 

After its formation, the planetesimal from Fig \ref{fig:Processing_LPC} was scattered inward by the giant planets, being thrust interior to Jupiter's orbit at $t \approx$600 kyr, before being scattered outward starting after $t\approx$800 kyr and eventually being implanted in the Oort cloud. The time that it spent under the dynamical control of Jupiter was an intense heating period that lasted $\sim$200 kyr, characterized by a very low effective thermal radius. Five short-period approaches can be easily identified during this period (upper panel of Figure \ref{fig:Processing_LPC}), associated with intense heating episodes in the lower panel of Figure \ref{fig:Processing_LPC}. While CO appears to sublimate continuously within 30 au, these approaches provoked the sublimation of \dioxi ice and the crystallization of amorphous water ice from a substantial subsurface layer of $\sim$90 m and $\sim$210 m respectively. It is important to clarify that these drops in the effective thermal radius and the ensuing heating episodes should not to be confused with the increased heating sustained during perihelion passages. The drops in effective thermal radius represent the planetesimal's tendency to orbit closer to the Sun over periods of time that are much longer than a typical cometary orbital period, comparable to the \textit{N}-body simulations time step of 100 and 1000 yrs. As the scattering toward the Oort Cloud begun, the planetesimal started to cool down, putting an end to the progression of the \dioxi and amorphous \water ice interfaces. Following the heat diffusion in deeper parts of its interior the CO interface continued to regress for some time, before becoming almost stagnant as the planetesimal moved beyond 100 au, and the simulation ended.

\subsection{Processing of populations}

Having presented an example of the coupling between the evolution models and the \textit{N}-body simulations, we now focus our attention to the three studied populations, searching for statistically significant differences that may characterize their degree of processing. The level of processing of each planetesimal, and therefore of each population, was tracked by the evolution of the ice content, as described by the retreat of the ice interfaces.

\subsubsection{Hyper-volatile content} \label{SubSec:Hyper_Volatiles}

\begin{figure}
    \centering
    \includegraphics[width=\linewidth]{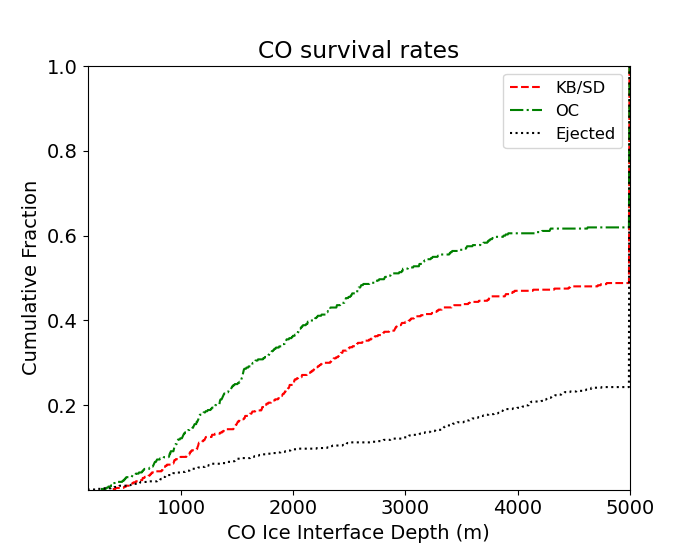}
    \caption{Cumulative distribution of the CO ice interface depths for the members of the three studied populations: \kb (red dashed line), \ooc (green dashed-dotted line) and ejected (black dotted line) planetesimals, at the end of their reservoir implantation and ejection phases respectively.}
\label{fig:CFD_CO}
\end{figure}

Figure \ref{fig:CFD_CO} shows the cumulative distribution of the depths of the CO ice interface at the end of the thermal evolution simulations for all the planetesimals of the ejected, \ooc, and \kb populations. A clear distinction in the degree of processing for the three populations is observed. Even though the ejected population had significantly shorter lifetimes in the Solar System, it is by far the most processed population. In $\sim$76\% of the ejected planetesimals the CO ice interface had already retreated to the center of their nuclei, before their ejection from the Solar System. The survival of pure CO ice close to the surface proved to be a highly unlikely event, with only $\sim$11\% of the planetesimals preserving some CO within 2500 m from the surface. 

Planetesimals implanted in the KB/SD were also substantially processed. In $\sim$51\% of planetesimals, the CO interface receded to the center of their nuclei at the end of the simulations. However, $\sim$33\% of the \kb planetesimals retained some quantity of their initial pure CO ice in the upper half of their nucleus. 

The least processed population, and the one with the most chances of preserving an important amount of its pure condensed hyper-volatile content, was the \ooc population, the progenitors of LPCs. Although 38\% of planetesimals were unable to retain their initial CO ice content, the chances of its survival were significantly higher than for the other populations. The possibility of founding the surviving pure CO ice close to the surface was higher as $\sim$46\% of the OC planetesimals maintained the CO ice interface in the upper part of their nuclei (within 2500 m from the surface) at the end of the simulations. We should clarify at this point, that this finding does not imply that any future activity would be triggered from such depths. It simply shows that an important reserve of pure CO ice can survive closer to the surface, when compared to the other populations, and it can become an important activity driver later on, as it can gradually reach the surface, provided the necessary temperature and pressure conditions. In addition, one should take into account the particularities of the return of these objects to the inner Solar System, which is quite different between the two populations -slow in the case of \kb objects vs sharp in the case of \ooc objects-, and is expected to increase these differences between the two populations \citep[e.g.][]{Gkotsinas_2022, Kaib_2022}.

In addition, we should remind at this point that our thermal evolution model did not explicitly account for a number of internal processes, by simultaneously resolving the heat and gas diffusion equations \citep[as done by][for example]{Davidsson_2021,Parhi_2023}. Instead, it relied on a simplified thermal evolution calculation coupled asynchronously to an ice evolution model. The latter assumed that sublimated gases were vented at the surface, similarly to some previous studies \citep{Choi_2002, Steckloff_2021}. However, recondensation both above and below an ice interface is expected if the appropriate temperature and pressure conditions are met \citep[see][in particular]{Parhi_2023}. With that in mind, we must clarify that the term `loss' for a given volatile, does not signify necessarily its complete depletion. It would be more precise to say that it describes a modification of the initial mass fraction because of thermal processing. Additional seasonal or diurnal cycles, acting on shorter timescales, should allow for the local depletion or enrichment of subsurface layers.

\subsubsection{Moderate and Low Volatility Content}

\begin{figure}
    \centering
    \includegraphics[width=\linewidth]{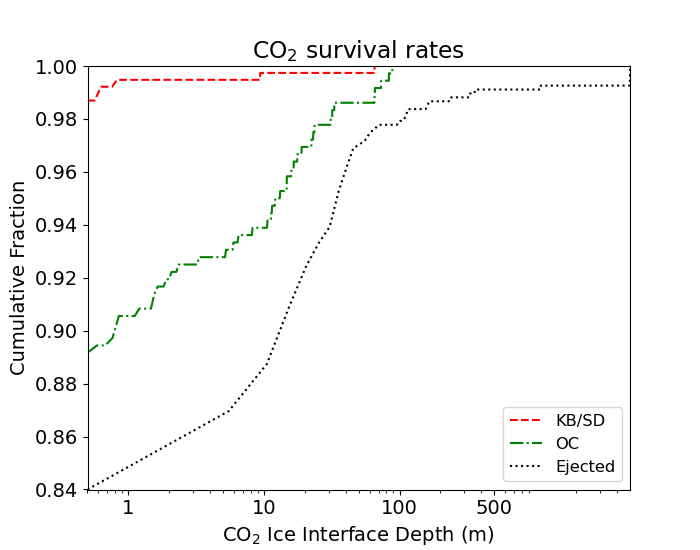}
    \caption{Cumulative distribution of the \dioxi ice interface depths for the members of the three studied populations: \kb (red dashed line), \ooc (green dashed-dotted line) and ejected (black dotted line) planetesimals, at the end of their reservoir implantation and ejection phases respectively.}
\label{fig:CFD_CO2}
\end{figure}

\begin{figure}
    \centering
    \includegraphics[width=\linewidth]{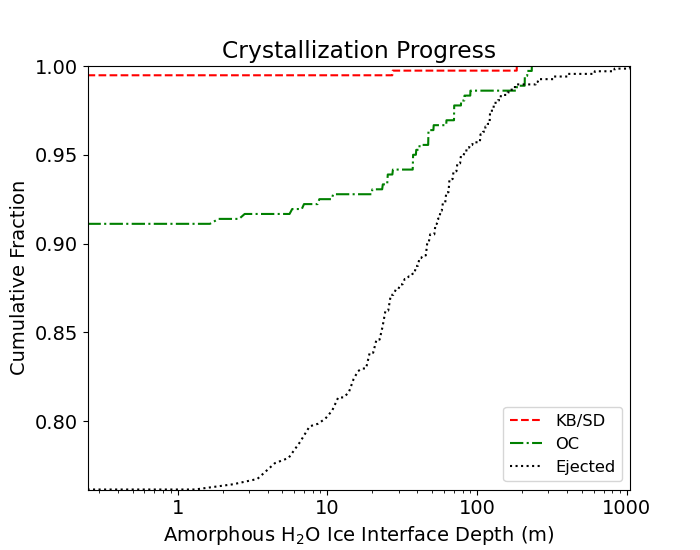}
    \caption{Cumulative distribution of the amorphous water ice crystallization interface depths for the members of the three studied populations: \kb (red dashed line), \ooc (green dashed-dotted line) and ejected (black dotted line) planetesimals, at the end of their reservoir implantation and ejection phases respectively.}
\label{fig:CFD_AM}
\end{figure}

Figures \ref{fig:CFD_CO2} and \ref{fig:CFD_AM} show the cumulative distribution of the depths of \dioxi and amorphous \water ice at the end of the thermal evolution simulations, similar to Figure \ref{fig:CFD_CO}. Compared to alterations on the hyper-volatile content, here, only a small fraction of planetesimals from each population undergoes the high temperature processing leading to the sublimation of \dioxi ice and the crystallization of amorphous \water ice. Nonetheless, in terms of \dioxi and amorphous \water ice, ejected planetesimals are again more processed on average than the `surviving' planetesimals, implanted in the \kb and the \ooc. 18\% of ejected planetesimals lost part of their \dioxi ice content, and in $\sim$25\%, the amorphous \water ice crystallized in a subsurface layer that did not exceed $\sim$100~m. Only a tiny fraction ($\sim$2\%) of ejected planetesimals underwent extensive processing, with ice interfaces receding as deep as $\sim$1000 m.

The affected planetesimals in the \kb and the \ooc populations were significantly lower: only $\sim$11\% and $\sim$9\% in the case of the Oort Cloud (for \dioxi and amorphous \water ice respectively) and just $\sim$1.5\% for the KB/SD planetesimals. These reduced fractions of affected planetesimals in the \ooc and \kb populations are a natural consequence on the one hand of the higher temperatures needed for these processes to take place and on the other hand of the scattering procedure that led them to the source reservoirs, as we briefly showed in Figure \ref{fig:OC_orbital_evolution} and as we explain in the next paragraphs. An overview of our simulations' results can be found in Table \ref{Table:Depths Early Processing}.

\subsubsection{The Imprint of the Orbital Evolution to the Thermal Processing}

As shown in Figure \ref{fig:OC_orbital_evolution}, \kb planetesimals were typically scattered outward by Neptune and Uranus \citep[e.g.][]{Nesvorny_2015}.  This happens over relatively long timescales, spending considerable time in the KB and the inner parts of the SD (with a$<$100 au) where temperatures are sufficient to trigger the sublimation of CO. However, this same pattern of orbital evolution resulted in the depletion of their condensed hyper-volatile content that protected their moderate- and low-volatility species. As they were mostly under the influence of the migrating Neptune, these planetesimals rarely roamed the area within Neptune's or Uranus' orbits, where temperatures are high enough for the sublimation of \dioxi ice and the crystallization of amorphous \water ice, before their implantation on stable orbits in the KB/SD.

Planetesimals implanted in the OC were mainly scattered outward by Jupiter and Saturn \citep[e.g.][]{Dones_2015,Kaib_Volk_2022}. This difference implies more passages in the inner parts of the giant planet area, and orbits closer to the Sun than the \kb population. These passages can eventually lead to intense, generally short-lived, heating episodes that can affect species of lower volatility, affecting a shallow subsurface layer, as in the example of Figure \ref{fig:Processing_LPC}. This also explains why if the \dioxi and crystallization interfaces were used as processing indicators, the \kb population appears to be less processed than the \ooc one, unlike the estimation based on the processing of CO. 

The same pattern seems to apply for the ejected planetesimals, which were for the most part scattered outward by Jupiter. Although their lifetimes in the Solar System were much shorter than that of \ooc planetesimals (see Section \ref{SubSec:Ending Conditions}), they spend it almost entirely in the inner parts of the Solar System, within the orbit of Jupiter, where they were exposed to much higher temperatures, leading to extensive processing, affecting all the primordial pure ice inventories.

\begin{deluxetable*}{cccccc}
\tablecaption{Final depths of the examined ice interfaces: CO, CO\textsubscript{2} and amorphous H\textsubscript{2}O ice, at the end of the simulations for the three populations: \kb, \ooc and ejected planetesimals.\label{Table:Depths Early Processing}}
\tablehead{
\colhead{Ice Species} & \colhead{Max (m)} & \colhead{Min (m)} & \colhead{Med. (m)} & \colhead{Mean (m)} & \colhead{$\sigma$ (m)}
}
\startdata
\multicolumn{6}{c}{\textbf{Kuiper-Belt/Scattered Disk}} \\
\hline
CO            & 5000.0 & 392.5 & 5000.0 & 3577.9 & 1619.6 \\
\dioxi        &   65.4 &   0.5 &    0.5 &    0.7 &    3.3 \\
H$_2$O$_{am}$ &  184.5 &   0.3 &    0.3 &    0.8 &    9.5 \\
\hline
\multicolumn{6}{c}{\textbf{Oort Cloud}} \\
\hline
CO            & 5000.0 & 287.1  & 2816.7 & 3081.9 & 1688.3 \\
\dioxi        &   88.5 &   0.5  &    0.5 &    2.5 &    9.7 \\
H$_2$O$_{am}$ &  232.5 &   0.3  &    0.3 &    6.1 &   27.3 \\
\hline
\multicolumn{6}{c}{\textbf{Ejected}} \\
\hline
CO            & 5000.0 & 167.4  & 5000.0 & 4420.6 & 1219.2 \\
\dioxi        & 5000.0 &   0.5  &    0.5 &   45.2 &  430.9 \\
H$_2$O$_{am}$ & 1052.5 &   0.3  &    0.3 &   16.5 &   66.6 \\
\enddata
\end{deluxetable*}

\section{Discussion} \label{Sec:Discusssion}

\subsection{On the survival of CO ice in the KB/SD} \label{Discussion_KB/SD} 

Evaluating the long-term global loss of condensed hyper-volatile species in the progenitors of JFCs has been the subject of several studies \citep{De_Sanctis_2001, Choi_2002, Davidsson_2021, Steckloff_2021, Parhi_2023}. It requires models solving simultaneously both the heat and gas diffusion equations (see also Section \ref{sec:Methods}), and simulations integrated on long timescales to evaluate whether this sublimation process is completed. 
For example, \citet{De_Sanctis_2001} performed such simulations on large objects (R = 80 km) evolving on stable orbits, so most likely post-reservoir implantation, at distances between 39 and 43~au, to assess the influence of insolation and radiogenic heating. After 10 Myr of simulation, they found that $\sim$30\% of the initial CO content was lost. Based on the exponential decline of the sublimation rate, they assumed that CO would be able to survive at the innermost parts of the objects. 

Similar studies were conducted by \citet{Choi_2002} and \citet{Steckloff_2021}. They found that condensed CO should be completely lost, neglecting however gas diffusion and assuming simply that the sublimating gas would vent at the surface. \citet{Davidsson_2021} simulated the coupled heat and gas diffusion on timescales long enough to actually assess the loss of pure condensed hyper-volatile species. They found that JFC-sized objects (R = 4 km) should lose all CO within 0.1~Myr, unless CO is trapped in some other medium such as amorphous \water ice or \dioxi ice. That timescale significantly increased with the objects' size, as 70 to 200 km objects were found to lose all condensed CO within 10 to 13 Myr. \citet{Parhi_2023} considered in their study more complex mixtures. They showed that all JFC-sized objects should lose all condensed hyper-volatile species, and that larger objects should not retain a primitive volatile mass ratio in their subsurface layers. Their work demonstrated an interesting sublimation and recondensation pattern, where volatile depletion proceeded from internal bulk sublimation, and from a sublimation front receding from the surface. A layer significantly enriched in hyper-volatiles developed just below the sublimation front, which was located close the surface. 

What this work adds to this picture is the orbital evolution of planetesimals. All studies mentioned above assumed icy objects on fixed orbits for the duration of their thermal simulations. Despite the different assumptions between the thermal evolution models and the considerations of the aforementioned analytical studies, our results are generally in agreement with the prediction of pure condensed CO ice depletion at heliocentric distances up to 45 au. However, beyond this distance, the chances for condensed CO ice survival increase.  We suggest an increasing survival rate for condensed CO for scattered disk objects and especially those placed at distances beyond 200 au, a region considered to be the main source for objects evolving into Jupiter-crossing orbits \citep[e.g.][]{Nesvorny_2015}. It is the consideration of their orbital evolution, by means of \textit{N}-body simulations, that revealed the effect of the chaotic nature of the trajectories on the thermal evolution of these planetesimals, bringing them back and forth from warmer to colder regions in a random way \citep{Gkotsinas_2022}. This proved to be a key factor in the retention of condensed CO ice in their interiors. Unlike previous studies, our objects underwent multiple heating and cooling, episodes during their implantation on the outer Solar System reservoirs, which acted as a significant `protection' mechanism for their buried, condensed hyper-volatile content.

\subsection{On the processing of Oort Cloud objects} 

The few studies that examined the thermal processing of objects in the Oort Cloud focused mainly on the effects of radioactive heating \citep{Haruyama_1993, Prialnik_1995}. Given the completely different set of initial assumptions, a comparison between their results and ours does not seem pertinent. A more prominent comparison can be made with some recent predictions by \citet{Davidsson_2021}, who noted that the increased timescales for CO loss in large objects (R = 70-200 Km), imply a dynamical evolution effect on the survival of CO, especially if they are scattered quickly toward the \ooc. This is what our results confirm, as the \ooc planetesimals proved to be the least processed population, a result of a relatively quick outward scattering. Indeed, the implantation on the \ooc took place on different timescales, and appeared to be a slow as much as a quick procedure (see Section \ref{SubSec:Ending Conditions}), leading to a variety of thermal processing outcomes, among which $\sim$60\% of the \ooc planetesimals managed to retain a part of their primordial pure condensed CO ice.
This is in disagreement with predictions for lack of hyper-volatile species in \ooc Comets, suggested by a recent analytical study \citep{Lisse_2022}. However, the differences in our results can be tracked back to dynamical timescales, as \citet{Lisse_2022} assumed that the implantation of planetesimals toward the \ooc starts after objects spent hundreds of Myr in the \kb region. Given that there is strong evidence that the giant planet instability took place no later than 100 million years after the start of Solar System formation \citep{Morbidelli_2018, Nesvorny_2018b, Mojzsis_2019}, and quite possibly within less than 10 million years \citep[e.g.][]{Clement_2018, Ribeiro_2020, Liu_2022}, this working assumption by \cite{Lisse_2022} seems rather unlikely.

Our prediction for condensed CO ice survival in the \ooc is consistent with the growing number of observations of suspected CO-driven activity on inbound LPCs, such as C/2017 K2 (PANSTARRS) \citep{Meech_2017}, C/2010 U3 (Boattini) \citep{Hui_2019} and C/2014 UN271 Bernardinelli-Bernstein \citep{Farnham_2021}. At the same time, the possibility of CO depletion or low CO abundance ratios, as observed for example on comet C/2021 A1 (Leonard) \citep{Faggi_2023} is not excluded at all. 

In addition, our results provide an explanation to the observed differences in the CO/CO$_2$ and CO/\water abundance ratios of LPCs \citep{A'Hearn_2012, Pinto_2022}, even if the only available data were obtained at short heliocentric distances (up to $\sim$6 au) and even if those are based on production rates which cannot be directly calculated by our model. According to our simulations these differences are an inherent consequence of the planetesimals' orbital history, an explanation that seems to be more plausible than a scenario for different formation areas for \kb and \ooc planetesimals, as suggested by \cite{A'Hearn_2012}. 

\begin{figure}
    \centering
    \includegraphics[width=\linewidth]{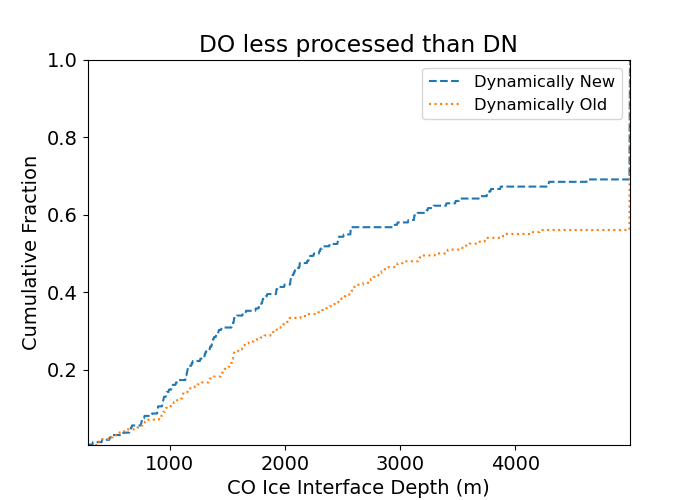}
    \caption{Cumulative distribution of the CO ice interface depths on planetesimals implanted at distances below the Oort spike ($a<$ 10000 au) and are considered to be Dynamically Old (DO) (orange dotted line), and planetesimals implanted beyond the Oort spike, considered to be Dynamically New (DN) (blue dashed line).}
\label{fig:DN_vs_DO}
\end{figure}

This explanation can be extended to the production rate differences between Dynamically New (DN) and Dynamically Old (DO) LPCs, observed initially by \cite{AHearn_1995} and confirmed recently by \cite{Pinto_2022}. Both studies observed that DN LPCs produce more \dioxi than CO, while DO LPCs appear to be more CO-dominant. This unexpected observation can be explained by the different trajectories leading to implantation on the OC. Assuming a typical value of 10000 au for the OC spike, separating DN to DO LPCs, we can see in Figure \ref{fig:DN_vs_DO}, that planetesimals which are statistically more likely to visit the planetary region for the first time (DN) are more CO-processed on average than planetesimals implanted below the Oort spike, an area which is considered to be the source of DO LPCs. This result does no exclude the possibility of further processing in the \ooc by galactic cosmic rays, as it has been previously proposed to explain this peculiar observation \citep{Pinto_2022}, but provides an additional and more concrete explanation, based on the implantation specifications on the \ooc.

We should note at this point, that this explanation does not automatically apply to comparisons between JFCs and LPCs as someone might expect. If for LPCs, our assumption seems more reliable given the abrupt nature of their return to inner Solar System, we cannot say the same thing for JFCs, that follow much more complicated and long trajectories toward the planetary region that is expected to provoke additional thermal processing \citep[e.g.][]{Gkotsinas_2022}, affecting further the chemical and physical properties of the comets. To make such comparisons, further simulations accounting for processing during the return to the inner Solar System are necessary, accompanied with stronger observational constraints on CO and \dioxi production rates, which for the time being are limited at distances within 3.5 au \citep{Pinto_2022}.      

\subsection{On drivers for cometary activity}

In agreement with these previous studies (see Section \ref{Discussion_KB/SD}), no significant alterations on the moderate (e.g. \dioxi or amorphous \water ices) and low-volatility (e.g. \water ice) content were recorded on planetesimals ending up in the \kb, when the only energy source considered is the solar irradiation. Alterations on the entire icy inventory, if any, should be the result of heating due to the decay of radioactive elements, affecting more significantly large objects (R$\geq$40 km) with low effective thermal conductivity \citep[e.g.][]{De_Sanctis_2001, Choi_2002, Malamud_2022}.
We found that for most \kb ($\sim$98\%) and \ooc objects ($\sim$90\%), \water ice remained in its initial amorphous phase during their implantation on their respective reservoirs, meaning that until their return to the inner solar system as JFCs and LPCs, it is highly unlikely to undergo any further alterations. 
This result implies different activity drivers upon the return of these objects in the inner Solar System as JFCs or LPCs. As CO seems to be more abundant and closer to the surface on \ooc objects, we would expect that activity, at least in large heliocentric distances, to be driven by CO sublimation. This prediction is in line with a growing number of suspected CO-driven activity on inbound LPCs \citep[e.g][]{Meech_2017, Hui_2019, Farnham_2021}. On the other hand, CO ice although present, seems to be less abundant on \kb objects, while the \dioxi and amorphous \water ice content appeared to remain almost intact and much closer to the surface. This suggests a larger diversity of potential drivers upon the return of \kb objects on the inner Solar System, with less pronounced CO-driven activity.

We would like to clarify here, that the results presented above and the ensuing interpretations should not be viewed as an attempt to explain a vast and very diverse body of observations regarding the activity of comets. This would require a more complicated evolution model, and even then, given the large number of free or unknown parameters, different outcomes should be expected. Instead, we aim to highlight the importance of the orbital evolution, a key aspect, so far disregarded, that is intimately linked to the thermally-induced physical and chemical evolution of cometary nuclei. By doing so, we showed that a variety of contradictory results, such as the predictions for CO depletion \citep[e.g.][]{Davidsson_2021, Lisse_2022, Parhi_2023} or the observation of long-distance CO-driven activity \citep[e.g.][]{Meech_2017}, can be explained in a very natural way. In addition, and despite the simplifications and the working assumptions, we showed that we can clearly identify statistically distinct behaviors between the different populations, even though they were formed and evolved from the same region of the protoplanetary disk.

\subsection{A critical review of our method} \label{Section:Critical review of our results}

In Section \ref{sec:Methods}, a novel approach was proposed to examine the thermal processing of planetesimals, by taking into account their long-term orbital evolution. In this approach, a number of simplifying assumptions were employed, rendering the coupling of the orbital to the thermal evolution feasible. In the next paragraphs we would like to examine the effect of these simplifications on the outcomes of our simulations, to better evaluate them. 

\subsubsection{Implications of Orbital Averaging}

One of the most prominent assumptions of our method is the use of equivalent average equilibrium temperature circular orbits. Although it proved to be a powerful tool from a numerical point of view, solving for instance several issues related to the output cadence of \textit{N}-body simulations (see also \cite{Gkotsinas_2023}), their use came with an important setback: They poorly accounted for high-temperature processes such as the sublimation of \water ice. 

As explained in \cite{Gkotsinas_2023}, all averaging techniques systematically place objects closer to their aphelion distance, resulting in the under-estimation of the surface temperature, especially close to perihelion. As a consequence, the simulated surface temperatures were insufficient for the onset of \water ice sublimation, and in some cases of \dioxi sublimation or the crystallization of amorphous \water ice.  

The \water ice sublimation from the surface and the associated erosion, should bring the buried ice interfaces closer to the surface, prompting an enhancement of the different processes as the incoming solar energy reaches these interfaces more easily. Unfortunately, such an effect is not accounted for in this study. However, given the low sublimation temperature of condensed CO, our results for this ice are not substantially affected, as the \dioxi and amorphous \water ice interfaces might be. With that in mind, our results for \ooc planetesimals that had prolonged residence times in the inner planetary region should be considered rather conservative on moderate and low-volatile species.

\subsubsection{Implications of the Thermal Modeling Assumptions}

In Sections \ref{Subsection:Thermal evolution model} and \ref{SubSec:Hyper_Volatiles} we described some of the impacts that our working assumptions for the thermal evolution model may have on our results. We highlighted in particular the short-term processes (of the order of minutes or hours) that are lacking in our methods. Indeed, ice recondensation above and below an ice interface should trigger a substantial redistribution of the ice species in a cometary interior. Moreover, we pointed out that our model examines the theoretically maximum processing of an area at the sub-solar point, receiving an averaged amount of solar energy that is decided from Equation \ref{Equation:Equilibrium_temperature}, meaning that no amplitude effects from any temporal or geographical variations were considered.

In Section \ref{SubSec:Asynchronous Model}, we discussed in length the effect of an accurate estimation of the saturation vapor pressure on CO ice conservation. Using classical formulas from \cite{Fanale_1990} (see also \cite{Prialnik_2004}), the chances of CO survival, even in the deepest part of the nucleus were close to zero. With more accurate estimations, such as the ones from \cite{Brown_1980, Fray_2009} and \cite{Grundy_2023}, we found that CO ice survival is instead possible. This observation points to the necessity of obtaining better quality constraints on parameters involved in cometary thermal evolution models (either from laboratory experiments or through observations), as slight changes in their values can radically change the outcomes of a given simulation. For example, some studies model the dust component as large aggregates of millimeter size \citep[e.g.][]{Gundlach_2012, Malamud_2015}, implying also larger pores, which are usually modelled to have a radius of the same order of magnitude as the grains \citep[e.g][]{Huebner_2006}. Here, the dust component was modelled as a powder with grain sizes of the order of 1 $\mu$m, a typical working assumption on many thermal evolution models \citep[e.g][]{Enzian_1997, Huebner_2006, Loveless_2022}. We tested the impact of such an assumption on a smaller sample of \kb planetesimals, using larger pores of the order of millimeter size to test the other end-member scenario. As expected, the effects on the diffusivity and subsequently on the ice loss (see Equations \ref{Equation:Local Diffusion Coef} and \ref{Eqution:Retreat_rate}) were severe, with a significant subsurface layer of the order of $\sim$2 km losing very quickly its CO content, before the procedure continues at a normal pace as seen in Figure \ref{fig:Processing_LPC}. This change in our initial parameters reduced greatly the chances of CO survival, for deep subsurfaces near the core of the nuclei, even for planetesimals we had identified as being able to maintain some CO in the other set of simulations. We note that other studies \citep[e.g.][]{Sarid_2009} employ a pore size distribution, often with thinner tubes near the surface, and wider in the interior of comets \citep{Prialnik_2017}, which seems to be a more appropriate approach.

\subsubsection{Implications of the Instability Timing} \label{SubSec:Instability_Timing_Implications}

An important parameter of this study is the timing of the giant planet instability, which marks the beginning of our simulations. While the occurrence of the instability is widely accepted because it can explain a number of features of the Solar System~\citep[see][for a review]{Nesvorny_2018}, the timing remains a matter of debate. The original 'Nice model' proposed an instability that was triggered relatively late ($\sim$500 Myr after the start of planet formation) to coincide with the `terminal lunar cataclysm' \citep{Gomes_2005, Morbidelli_2012}. Recent works have shown that the inferred cataclysm almost certainly was the result of sampling bias~\citep{Boehnke_2016, Zellner_2017, Hartmann_2019}. In addition, a late giant planet instability almost invariably would have destabilized the orbits of the terrestrial planets, assuming them to be fully-formed at that time \citep{Kaib_2016}. Various cosmochemical constraints agree that the instability must have taken place earlier than 100 million years after the start of planet formation \citep{Morbidelli_2018, Nesvorny_2018b, Mojzsis_2019}. From a dynamical point of view, the instability could have been triggered by the dissipation of the gaseous disk \citep{Liu_2022}, by planet-planet interactions after the dissipation of the disk \citep{Ribeiro_2020}, or by interactions between the giant planets and the primordial planetesimal disk \citep[the 'classic' instability trigger;][]{Tsiganis_2005, Quarles_2019, Ribeiro_2020}. A gas disk-driven instability would have taken place on the disk dissipation timescale of a few Myr \citep{Haisch_2001}. A self-driven instability would also have taken place quickly, within $\sim$3-5 Myr of gas disk dispersal \citep{Ribeiro_2020}. A planetesimal disk-driven instability, however, would have taken $\sim$50 Myr to be triggered \citep{Quarles_2019, Ribeiro_2020}. A gas disk-driven or self-driven instability would have taken place while the terrestrial planets were still forming \citep{Clement_2018}, whereas a planetesimal disk-driven instability would have taken place after the main formation phase. In any case, the time between the gas disk dispersal and the triggering of the giant planet instability represents a calm period of thermal processing before the planetesimals were scattered toward their reservoirs or ejected from the Solar System. As discussed above, the main difference between our results and those of \cite{Lisse_2022} -- who predict a lack of hyper-volatiles in \ooc comets -- comes from \cite{Lisse_2022}'s assumption of a several hundred million year-long stable processing phase, which is at odds with the current interpretation of early Solar System evolution. 

\citet{Davidsson_2021} examined the processing on a scenario where the instability is triggered after 15 Myr from the formation of the disk. By considering objects evolving at distances of 23 au on circular orbits, they suggested that all JFC-sized planetesimals (R = 4 km) should lose their condensed CO content in $\sim$0.1~Myr. Their results indicate that a prolonged stay in the disk would result in significant thermal processing, pointing to the necessity of establishing a precise timeline for the instability, to get a better picture of the early processing of future comets. More importantly, understanding the timing of planetesimal formation would allow to more appropriately account for the effect of internal radiogenic heating, which we have neglected in this study \citep[e.g.][]{Prialnik_1987, Haruyama_1993, Prialnik_1995, Prialnik_2006, Malamud_2022}. 

\section{Summary} \label{Sec: Conclusions}

We modeled the thermal evolution of 1518 planetesimals originating from the planetesimal disk and populating outer Solar System reservoirs or escaping from the Solar System to the interstellar space. To this end, we used a dedicated thermal and ice evolution model coupled to planetesimals' trajectories, obtained from two sets of \textit{N}-body simulations. Our simulations showed that:

\begin{itemize}
\item Ejected planetesimals were the most processed, with \kb planetesimals being significantly less altered, and \ooc planetesimals being the least processed population on average.

\item The thermal processing during the implantation to the \kb and the \ooc or the ejection from the Solar System affected primarily the pure condensed CO ice content. Less volatile species like \dioxi or amorphous \water ice did not present important alterations on the majority of the planetesimals.

\item Unlike previous predictions, the survival of CO ice, and therefore of other species of high volatility, is possible in a significant number of \kb and \ooc objects. This provides a natural explanation for the observed variability in the abundance ratios of CO on JFC and LPC populations, and a framework for distinct drivers of cometary activity for these populations upon their return as JFCs and LPCs.

\item An intimate relation between the orbital and the thermal evolution of planetesimals was established, pointing to the necessity of taking into account the former to obtain the full picture of any past alterations on comets observed in the current epoch. 
\end{itemize}

\begin{acknowledgments}
We would like to thank the two anonymous reviewers for comments and suggestions that greatly improved this article. We are grateful to Will Grundy for sharing CO data ahead of their publication. We also thank Emmanuel Lellouch and Matthew M. Knight, who reviewed this work as part of AG's PhD defense, and made comments that greatly improved it. This study is part of a project that has received funding from the European Research Council (ERC) under the European Union’s Horizon 2020 research and innovation program (Grant agreement No. 802699). We gratefully acknowledge support from the PSMN (Pôle Scientifique de Modélisation Numérique) of the ENS de Lyon for the computing resources. S.N.R. is grateful to the CNRS's MITI and PNP programs, and also acknowledges funding from the French Programme National de Planétologie (PNP) and in the framework of the Investments for the Future programme IdEx, Université de Bordeaux/RRI ORIGINS.
\end{acknowledgments}

\appendix

\section{Physical Parameters of the Thermal Evolution Model} \label{Sec:Appendix}

The effective thermal conductivity accounts for reduction entailed by the porous nature of cometary material. Here we use Russell's correction factor \citep{Russell_1935}, a widely used expression in models for the thermal evolution of comets \citep[e.g.][just to name a few]{Espinasse_1991, Orosei_1999}:
\begin{equation}
  \phi = \frac{\psi^{2/3} f + (1-\psi^{2/3})}{\psi - \psi^{2/3} + 1 - \psi^{2/3}(\psi^{1/3} - 1)f},
\end{equation}
\noindent where $f$ is the ratio of the solid to the radiative conductivity, which can be calculated as:
\begin{equation}
   \kappa_{rad} = 4 \epsilon \sigma r_p T^3 \label{Equation:Conductivity_Radiation}
\end{equation}
\noindent with $\epsilon$ the material's emissivity, $\sigma$ (W m\textsuperscript{-2} K\textsuperscript{-4}) the Stefan-Boltzmann constant, $r_p$ (m) the radius of the pores and $T$ (K) the temperature of the matrix.

\begin{deluxetable*}{cccccc}
\tablecaption{Physical parameters of the nominal object.\label{Table:Physical Parameters}}
\tablehead{
\colhead{Parameter} & \colhead{Symbol} & \colhead{Value} & \colhead{Units} & \colhead{Reference}
}
\startdata
Density      & $\rho_{CO}$          & 890             & kg m$^{-3}$         & \cite{Luna_2022}      \\
\nodata      & $\rho_{CO_2}$        & 1500            & \nodata         & \cite{Satorre_2008}   \\
\nodata      & $\rho_{H_2O_{cr}}$   & 917             & \nodata         & \cite{Eisenberg_2005} \\
\nodata      & $\rho_{H_2O_{am}}$   & 940             & \nodata         & \cite{Ghormley_1971}  \\
\nodata      & $\rho_{dust}$        & 3000            & \nodata         & \cite{Enzian_1997}    \\
Heat Capacity& c$_{CO}$             & 35.7$T$ - 187   & J kg$^{-1}$K$^{-1}$ & \cite{Clayton_1932}   \\
\nodata      & c$_{CO2}$            & 6.34$T$ + 167.8 & \nodata & \cite{Giauque_1937}   \\
\nodata      & c$_{H_2O}$           & 7.49$T$ + 90    & \nodata & \cite{Giauque_1936}   \\
\nodata      & c$_{dust}$           & 3$T$            & \nodata & \cite{Enzian_1997}    \\
Conductivity & $\kappa_{CO}$        & 0.1             & W m$^{-1}$K$^{-1}$  & \cite{Munoz_Caro_2016}\\
\nodata      & $\kappa_{CO_2}$      & 93.4/$T$        & \nodata  & \cite{Mellon_1996}      \\
\nodata      & $\kappa_{H_2O_{am}}$ & 2.34$\cdot$10$^{-3}$T+2.8$\cdot$10$^{-2}$ & \nodata & \cite{Klinger_1980} \\
\nodata      & $\kappa_{H_2O_{cr}}$ & 567/$T$         & \nodata & \cite{Klinger_1980}     \\
\nodata      & $\kappa_{Dust}$      & 2.5             & \nodata & -                       \\
Bond's Albedo & $\mathcal{A}$       & 0.04            & -                  & -                       \\
IR Emissivity & $\varepsilon$       & 0.96            & -                  & -                       \\
Hertz factor  & $h$                 & 0.01            & -                  & \cite{Huebner_2006}     \\
Mean Pore Radius & $r_p$            & 10$^{-6}$       & m                  & -                       \\
Tortuosity       & $\xi$            & 1               & -                  & -                       \\
Local zenith angle & $\zeta$        & 0               & rad                & -                       \\
Porosity           & $\psi$         & 0.75            & -                  & \cite{Jorda_2016}       \\
Nominal radius     & R              & 5000            & m                  & -                       \\
Initial Temperature& T              & 10              & K                  & -                       \\
\enddata
\end{deluxetable*}

\begin{deluxetable*}{cr} \label{Table:Saturation Pressure Coeff}
\tablecaption{Coefficients for the saturation vapor pressure of CO, given by Equation \ref{Eq:Saturation Vapor CO}}
\tablehead{
\colhead{Coefficient} & \colhead{Value}
}
\startdata
$A_0$ & 1.80741183$\cdot$10$^{1}$  \\
$A_1$ & -7.69842078$\cdot$10$^{2}$ \\
$A_2$ & -1.21487759$\cdot$10$^{4}$ \\
$A_3$ & 2.73500950$\cdot$10$^{5}$  \\
$A_4$ & -2.90874670$\cdot$10$^{6}$ \\
$A_5$ & 1.20319418$\cdot$10$^{7}$  \\
\enddata
\end{deluxetable*}

\bibliography{references}{}
\bibliographystyle{aasjournal}

\end{document}